\documentclass[12pt]{article}
\pdfoutput=1
\usepackage{graphicx}
\usepackage{epsfig}
\def\beq{\begin{equation}}
\def\eeq{\end{equation}}
\def\bea{\begin{eqnarray}}
\def\eea{\end{eqnarray}}
\def\nn{\nonumber}

\def\roughly#1{\mathrel{\raise.3ex\hbox
{$#1$\kern-.75em\lower1ex\hbox{$\sim$}}}}

\def\sla#1{\raise.15ex\hbox{$/$}\kern-.57em #1}

\long\def\symbolfootnote[#1]#2{\begingroup%
\def\thefootnote{\fnsymbol{footnote}}\footnote[#1]{#2}\endgroup}

\begin{document}

\begin{flushright}
UMiss-HEP-2008-05  \\
UdeM-GPP-TH-07-171 \\
\end{flushright}

\begin{center}
\bigskip
{\Large \bf \boldmath CP Violation in $\tau\to K\pi\pi
\nu_\tau$} \\
\bigskip
\bigskip
{\large 
Ken Kiers $^{a,}$\symbolfootnote[1]{knkiers@taylor.edu},
Kevin Little $^{a,}$\symbolfootnote[2]{little@uchicago.edu.  
(Address after September 15, 2008:
Department of Radiology, MC2026,
University of Chicago,
5841 South Maryland Avenue,
Chicago, IL 60637, USA.)}
Alakabha Datta $^{b,}$\symbolfootnote[3]{datta@phy.olemiss.edu},
David London $^{c,}$\symbolfootnote[4]{london@lps.umontreal.ca}, \\
Makiko Nagashima $^{c,}$\symbolfootnote[5]{makiko@lps.umontreal.ca},
and Alejandro Szynkman
$^{c,}$\symbolfootnote[6]{szynkman@lps.umontreal.ca} }
\end{center}

\begin{flushleft}
~~~~~~~~~~~$a$: {\it Physics Department, Taylor University,}\\
~~~~~~~~~~~~~~~{\it 236 West Reade Ave., Upland, Indiana, 46989, USA}\\
~~~~~~~~~~~$b$: {\it Dept of Physics and Astronomy, 108 Lewis Hall,}\\
~~~~~~~~~~~~~~~{\it University of Mississippi, Oxford, MS 38677-1848, USA}\\
~~~~~~~~~~~$c$: {\it Physique des Particules, Universit\'e
de Montr\'eal,}\\
~~~~~~~~~~~~~~~{\it C.P. 6128, succ. centre-ville, Montr\'eal, QC,
Canada H3C 3J7}\\
\end{flushleft}

\begin{center}
\bigskip (\today)
\vskip0.5cm {\Large Abstract\\} \vskip3truemm
\parbox[t]{\textwidth}{We consider CP-violating effects in $\tau\to K
  \pi\pi\nu_\tau$, assuming that a charged Higgs boson provides a new
  amplitude that can interfere with the usual Standard Model
  amplitude.  We consider four CP-odd observables -- the regular rate
  asymmetry, two modified rate asymmetries and a triple-product
  asymmetry.  The regular rate asymmetry is expected to be small
  because it requires the interference of the new physics amplitude
  with the standard model amplitude containing the hadronic scalar
  form factor.  The other CP asymmetries may be more promising in
  terms of their new physics reach.  Numerical estimates indicate that
  the maximum obtainable values for the modified and triple-product
  asymmetries are on the order of a percent.  }
\end{center}

\thispagestyle{empty}
\newpage
\setcounter{page}{1}
\baselineskip=14pt

\setcounter{footnote}{0}

\section{Introduction}

In the standard model (SM) of particle physics, CP violation is due to
a complex phase in the Cabibbo-Kobayashi-Maskawa (CKM) matrix. But is
this the only source of CP violation? In order to answer this
question, it is important to look for CP-violating effects in as many
systems as possible.

One such system is $\tau$ decays. In the SM, CP violation in the
$\tau$ system is essentially zero~\cite{BS}; we consider, instead, a
search for physics beyond the SM. In Ref.~\cite{DeltaS=0}, we examined
CP violation in strangeness-conserving $\tau$ decays. It is only
natural next to turn to those processes with $\Delta S =1$. The
simplest such decay is $\tau \to K \pi \nu_\tau$. However, CP
violation in this process has been analyzed in detail in
Ref.~\cite{KMPLB398}, and we have nothing to add here. The next decay
is $\tau \to K \pi \pi \nu_\tau$. This has been examined theoretically
in the past in Refs.~\cite{kilian,finkemeiermirkes2}. In this paper we
update these analyses~\cite{fpcp2008}.

One has to assume the presence of new physics in order to get non-zero
CP-violating effects when comparing $\tau^- \to K^- \pi^-
\pi^+\nu_\tau$ to its CP-conjugate decay.  In Ref.~\cite{kilian}, the
left-right (LR) model is assumed when the authors consider $\tau\to K
\pi \pi \nu_\tau$.  However, as shown in Ref.~\cite{DeltaS=0}, if
there is no LR mixing, CP violation is proportional to the mass of the
neutrino, and is negligible. Thus, in the LR model, CP violation in
$\tau$ decays is proportional to $W_L$-$W_R$ mixing. However, we know
this is small \cite{LRmixing}. We therefore conclude that sizeable CP
violation in the $\tau$ system cannot arise in the LR model.

For this reason, in this paper, we assume that the $\tau$ decay
includes the exchange of a new-physics (NP) charged Higgs. Note that
many NP models have two Higgs doublets, so that a charged Higgs is
present. However, if the Higgs doublets give mass to the fermions, the
coupling of the charged Higgs boson is generally proportional to the
masses of the first- and second-generation quarks. Since these are
small, CP violation in the $\tau$ system will also be small. To avoid
this, if CP violation is to be observed in $\tau$ decays, the
charged-Higgs coupling must be large. In other words, $\tau \to K \pi
\pi \nu_\tau$ probes non-``standard'' NP CP violation.

It is worth noting at this point that CLEO has searched for CP
violation in $\tau\to K\pi\nu_\tau$~\cite{CPVKpiexpt} and has set a
bound on a coupling constant related to the scalar coupling of a
charged Higgs (or other scalar boson) to the up and strange quarks.
The experimental investigation suggested in this work would be
complementary to that carried out in Ref.~\cite{CPVKpiexpt} in that it
would probe the pseudoscalar coupling of the Higgs to the up and
strange quarks.  In the notation introduced below, the CLEO experiment
probed $\eta_S$, while a CP analysis of $\tau\to K\pi\pi \nu_\tau$
would probe $\eta_P$ [see Eq.~(\ref{eq:np_hamiltonian}) below].

In the presence of one NP contribution, the amplitude for the decay
$\tau \to K \pi \pi \nu_\tau$ can be written
\beq{\cal A} = A_1 + A_2 e^{i\phi} e^{i\delta} ~,
\eeq
where $\phi$ and $\delta$ are the relative weak (CP-odd) and strong
(CP-even) phases, respectively. The amplitude for the antiprocess,
${\bar{\cal A}}$, is given by the same expression, but with $\phi \to
-\phi$.

In general, CP violation is obtained by comparing $|{\cal A}|^2$ to
$|{\bar{\cal A}}|^2$. There are three types of signals:
\begin{enumerate}

\item The full rate for a particular process involves
  $\sum_{\textrm{\scriptsize{spins}}} |{\cal A}|^2$, integrated over
  the final-state momenta in the usual way. The {\it rate asymmetry}
  is given by the rate difference of the process and antiprocess.

\item The rate asymmetry can be altered in two ways. First, if some
  spins are measured, one does not sum over them.  Alternatively, one
  can integrate asymmetrically in order to isolate certain terms in
  the differential width. In either case the process-antiprocess
  difference leads to a {\it modified rate asymmetry}.

\item One can also construct CP asymmetries based on the quantity
  $\vec v_1 \cdot (\vec v_2 \times \vec v_3)$, where each $v_i$ is a
  spin or momentum. This is a {\it triple product} (TP), and its value
  can be different for process and antiprocess, signaling CP
  violation.

\end{enumerate}
The rate asymmetry or modified rate asymmetry is proportional to
\beq 
\sin \delta \sin \phi 
\label{rateangles}
\eeq
(integrated over phase space).  Thus, this category of CP violation
requires that the two decay amplitudes have a non-zero relative weak
{\it and} strong phase. The TP asymmetry is proportional to
\beq 
\cos \delta \sin \phi ~,
\label{TPangles}
\eeq
so that one does not require a strong-phase difference to get a TP
asymmetry. In this paper we consider all three types of CP violation
in $\tau \to K \pi \pi \nu_\tau$.  (Ref.~\cite{kilian} considers only
TP's.)

The remainder of this paper is organized as follows.  In
Sec.~\ref{sec:diffwidth}, we write down the expression for the
differential width for $\tau^-\to K^-\pi^-\pi^+ \nu_\tau$ in terms of
various form factors, and including the NP contribution.  We perform
weighted integrations of the differential width over phase space to
isolate certain cross-terms.  In Sec.~\ref{sec:cpodd}, we consider
four CP asymmetries: the regular rate asymmetry, two modified rate
asymmetries, and a triple-product asymmetry.  The modified rate
asymmetries and triple-product asymmetry are constructed using the
weighted differential widths from Sec.~\ref{sec:diffwidth}.
Section~\ref{sec:numresults} contains a numerical analysis of the
modified rate asymmetries and the triple-product asymmetry.  We
conclude with a few closing remarks in Sec.~\ref{sec:conclusions}.

\section{Differential width for $\tau^-\to K^-\pi^-\pi^+
\nu_\tau$}
\label{sec:diffwidth}

We start by determining an expression for the differential width for
$\tau^-\to K^-\pi^-\pi^+\nu_\tau$, including possible NP effects due
to a new charged Higgs boson $H^-$.

\subsection{General expression for the differential width}

Let us begin by considering the SM contribution to $\tau^-\to
K^-\pi^-\pi^+\nu_\tau$. Within the SM, the relevant effective
Hamiltonian is given by
\bea
   {\cal H}^{\textrm{\scriptsize{SM}}}_{\textrm{\scriptsize{eff}}} = 
    \frac{G_F}{\sqrt{2}}\sin\theta_c \, 
      \bar{\nu}_\tau\gamma_\mu(1-\gamma_5)\tau 
    \, \bar{s} \gamma^\mu(1-\gamma_5) u + \mbox{h.c.,}
\label{eq:sm_hamiltonian}
\eea
where $\theta_c$ is the Cabibbo angle.  The hadronic matrix element
for the decay may be conveniently parameterized in terms of four form
factors as follows~\cite{KM1992},
\bea
   J^\mu & \equiv & \langle K^-(p_1) \pi^-(p_2) \pi^+(p_3)| \bar{s}\gamma^\mu
     (1-\gamma^5) u | 0\rangle \nn \\
     &=& \left[F_1(s_1,s_2,Q^2) (p_1-p_3)_\nu +
       F_2(s_1,s_2,Q^2) (p_2-p_3)_\nu \right]T^{\mu\nu} \nn \\
     && + i F_3(s_1,s_2,Q^2) \epsilon^{\mu\nu\rho\sigma}
         p_{1\nu}p_{2\rho}p_{3\sigma} + F_4(s_1,s_2,Q^2) Q^\mu\, ,
\label{eq:jmu_sm}
\eea
where $Q^\mu=(p_1+p_2+p_3)^\mu$, $T^{\mu\nu}=g^{\mu\nu}-Q^\mu
Q^\nu/Q^2$, $s_1=(p_2+p_3)^2$ and $s_2=(p_1+p_3)^2$; also, we adopt
the convention $\epsilon_{0123}=+1$ as in
Ref.~\cite{KM1992}.\footnote{The authors of Ref.~\cite{KM1992} adopt
  the convention $\epsilon_{0123}=+1$, but don't state the precise
  functional form for $F_3$.  Subsequent authors state $F_3$, but the
  sign of $\epsilon_{0123}$ is not obvious.  We make a particular
  choice for the sign of $F_3$ below; changing this sign would change
  the sign of the related asymmetry.}  The form factors $F_1$-$F_4$
have been considered, for example, in Ref.~\cite{dmsw}. As noted
there, $F_1$ can arise due to the decay chain $\tau\to K_1\nu_\tau$,
with $K_1\to K^*\pi\to K\pi\pi$, while $F_2$ comes from $\tau\to
K_1\nu_\tau$, with $K_1\to K \rho \to K\pi\pi$.  It is now known that
both the $K_1(1270)$ and the $K_1(1400)$ contribute (see
Sec.~\ref{sec:ff} for further details).  $F_3$ is the anomalous
Wess-Zumino term. It can be estimated by considering the decay chain
$\tau \to K^*\nu_\tau$, with the intermediate $K^*$ going to $\rho K$
or $K^* \pi$~\cite{dmsw}. The scalar term, $F_4$, is generally assumed
to be negligible for this decay, since there is no suitable
pseudoscalar resonance through which the decay can proceed.  The
authors of Ref.~\cite{dfm1994} performed a calculation of $F_4$ within
the context of Chiral Perturbation Theory and found that $F_4$ is
non-zero if one includes chiral-symmetry-breaking mass terms for the
quarks.  The resulting expression for $F_4$ was found to contain both
a non-resonant term (proportional to $m_\pi^2+m_K^2$) and a resonant
term.  A numerical study indicated that the SM scalar contribution to
the width was quite small~\cite{dfm1994}.  We will consider the form
factors further in Sec.~\ref{sec:ff}. At this point we simply note
that $F_1$ and $F_2$ give the dominant contributions to the rate for
$\tau\to K\pi\pi\nu_\tau$~\cite{CLEO_K1}, while numerical estimates
indicate that the Wess-Zumino term ($F_3$) gives a subdominant
contribution.  In fact, in their experimental analysis, CLEO discards
the Wess-Zumino term altogether and considers only the contributions
due to $F_1$ and $F_2$~\cite{CLEO_K1}.

Starting from Eq.~(\ref{eq:jmu_sm}), the amplitude squared
for $\tau^-\to K^-\pi^-\pi^+\nu_\tau$ within the context of
the SM is given by
\bea
    \left|{\cal A}_{\textrm{\scriptsize{SM}}}\right|^2 
      & = & \frac{G_F^2}{2}\sin^2\theta_c 
      L_{\mu\nu}H^{\mu\nu} \; ,
   \label{eq:Asq}
\eea
where $L_{\mu\nu}=M_\mu\left(M_\nu\right)^\dagger$ and
$H^{\mu\nu}=J^\mu\left(J^\nu\right)^\dagger$, with
$M_\mu=\bar{u}_{\nu_\tau}\gamma_\mu(1-\gamma^5)u_\tau$.

Effects due to a charged Higgs modify the effective Hamiltonian
relevant for $\tau^-\to K^-\pi^-\pi^+\nu_\tau$, adding the following
terms,\footnote{These expressions are similar to those in
  Ref.~\cite{kilian}, although our notation differs slightly from that
  found there.}
\bea
   {\cal H}^{\textrm{\scriptsize{NP}}}_{\textrm{\scriptsize{eff}}} =
    \frac{G_F}{\sqrt{2}}\sin\theta_c \left[
      \eta_{S} \bar{\nu}_\tau(1+\gamma_5)\tau 
    \, \bar{s}  u  +
      \eta_{P} \bar{\nu}_\tau(1+\gamma_5)\tau 
    \, \bar{s} \gamma_5 u \right] + \mbox{h.c.}
\label{eq:np_hamiltonian}
\eea 
The total effective Hamiltonian is then ${\cal
  H}_{\textrm{\scriptsize{eff}}}= {\cal
  H}^{\textrm{\scriptsize{SM}}}_{\textrm{\scriptsize{eff}}}+ {\cal
  H}^{\textrm{\scriptsize{NP}}}_{\textrm{\scriptsize{eff}}}$. In
writing down Eq.~(\ref{eq:np_hamiltonian}) we have neglected terms
that would involve a right-handed projection of the neutrino field.
The interference of such terms with the SM amplitude would be
suppressed by $m_{\nu_i}$ (assuming that the neutrino spin states are
summed over).

The NP effects can be incorporated into the amplitude in a
straightforward manner. We first define a new current
$\widetilde{J}^\mu$, which is obtained from $J^\mu$ by the replacement
\bea
   F_4 & \rightarrow & \widetilde{F}_4=F_4
     + \frac{f_H}{m_\tau} \eta_P \, ,
\label{eq:f4_mod}
\eea
where the pseudoscalar form factor has been defined as follows
\bea
   \langle K^-(p_1) \pi^-(p_2) \pi^+(p_3)| \bar{s}
     \gamma^5 u | 0\rangle = f_H.
\eea
Defining $\widetilde{H}^{\mu\nu}\equiv
\widetilde{J}^\mu\left(\widetilde{J}^\nu\right)^\dagger$, we then find
the following expression for the square of the matrix element,
\bea
    \left|{\cal A}\right|^2 & = & \frac{G_F^2}{2}\sin^2\!\theta_c 
     L_{\mu\nu}\widetilde{H}^{\mu\nu} \; .
   \label{eq:Asq_2}
\eea
Note that we have used the $\tau^-$ equation of motion in order to
arrive at our definition of $\widetilde{F}_4$.  We have also neglected
the mass of the neutrino.

\begin{figure}[t]
\begin{center}
\resizebox{4in}{!}{\includegraphics*{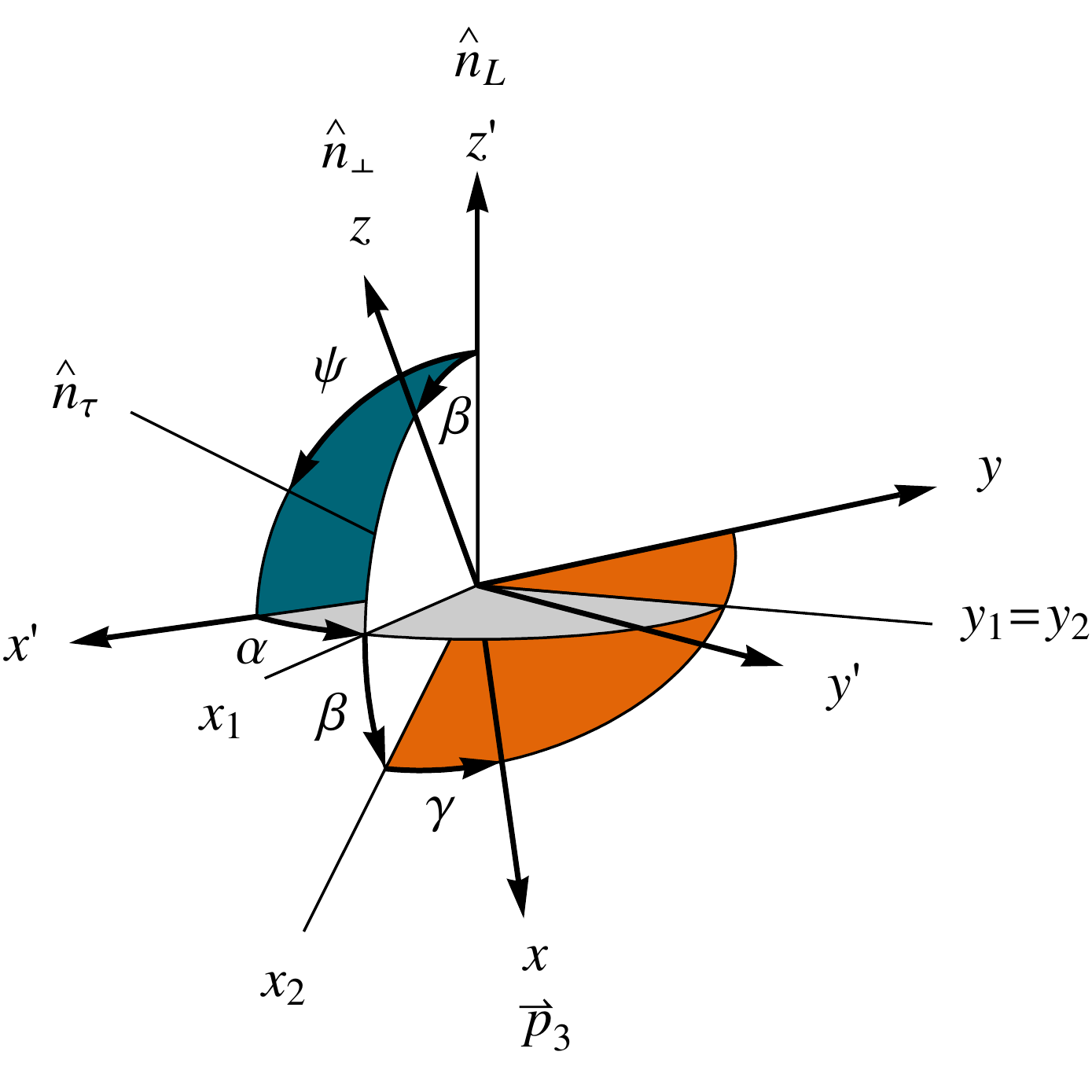}}
\caption{Definitions of the angles $\alpha$, $\beta$, $\gamma$ and
  $\psi$.  The Euler rotations corresponding to $\alpha$, $\beta$ and
  $\gamma$ are about the $z^\prime$, $y_1=y_2$ and $z$ axes,
  respectively.  This figure is very similar to a figure found in
  Ref.~\cite{KM1992}.  }
\label{fig:angles}
\end{center}
\end{figure}

The decays of $\tau$ leptons to final states containing two and three
pseudoscalar mesons have been thoroughly analyzed in
Ref.~\cite{KM1992}. The notation described there has been adopted
widely in the field and is quite standard. First, let us define
several useful angles. Our definitions are identical to those in
Ref.~\cite{KM1992}. We review the various definitions here for
convenience (more details may be found in Ref.~\cite{KM1992}).  The
angle $\theta$ is defined in the $\tau$ rest frame; in that frame it
is the angle between the direction of the hadrons (``$\vec{Q}$'') and
the direction of the tau in the laboratory frame. All other angles are
defined in the hadronic rest frame (i.e., the frame in which
$\vec{Q}\equiv\vec{p}_1+\vec{p}_2+\vec{p}_3=0$). In the hadronic rest
frame we define two different coordinate systems, $S$ and $S^\prime$.
These two coordinate systems are related by an Euler rotation using
the Euler angles $\alpha$, $\beta$ and $\gamma$, as indicated in
Fig.~\ref{fig:angles}. The $z^\prime$ axis in $S^\prime$ is chosen as
the direction of the laboratory in the hadronic rest frame
($\hat{n}_L$). The $x^\prime$ axis is chosen such that the $\tau$
direction ($\hat{n}_\tau$) is in the $x^\prime$-$z^\prime$ plane,
making an angle $\psi$ with respect to the $z^\prime$ axis (see
Fig.~\ref{fig:angles}). The $z$ axis in $S$ is perpendicular to the
plane defined by the momenta of the hadrons: $\hat{z} = \hat{n}_\perp
\equiv \vec{p}_1\times\vec{p}_2 /
\left|\vec{p}_1\times\vec{p}_2\right|$.  The $x$ axis is taken to be
the direction of $\vec{p}_3$; i.e., $\hat{x}=
\vec{p}_3/\left|\vec{p}_3\right|$. The three Euler angles are defined
as follows: $\alpha$ is the angle between the
$(\hat{n}_L,\hat{n}_\tau)$ plane and the $(\hat{n}_L,\hat{n}_\perp)$
plane, $\beta$ is the angle between $\hat{n}_L$ and $\hat{n}_\perp$
and $\gamma$ is the angle between the $(\hat{n}_L,\hat{n}_\perp)$
plane and the $(\hat{n}_\perp,\hat{x})$ plane.

%
%

Having defined the various angles, we may write the differential width
for $\tau^-\to K^-\pi^-\pi^+\nu_\tau$ as follows~\cite{KM1992},
\bea
   d\Gamma & = & 
     \frac{G_F^2\sin^2\!\theta_c}{256\left(2\pi\right)^5m_\tau}
     \frac{m_\tau^2-Q^2}{m_\tau^2}
     \frac{dQ^2}{Q^2}
     \,ds_1 \,ds_2\,\frac{d\alpha}{2\pi}
     \frac{d\gamma}{2\pi}
     \frac{d\cos\beta}{2}
     \frac{d\cos\theta}{2} L_{\mu\nu}\widetilde{H}^{\mu\nu} \, ,
\label{eq:dgam}
\eea
where $Q^2$, $s_1$ and $s_2$ were defined below Eq.~(\ref{eq:jmu_sm}).

The coordinate system $S$ is convenient for expressing the momenta of
the three pseudoscalar mesons and for computing the various components
of the tensor $\widetilde{H}^{\mu\nu}$. In this coordinate system we
have~\cite{KM1992},
\bea
   p_1^\mu & = & (E_1,p_1^x,p_1^y,0)\, , \\
   p_2^\mu & = & (E_2,p_2^x,p_2^y,0)\,, \\
   p_3^\mu & = & (E_3,p_3^x,0,0)\,, 
\eea
where
\bea
   E_i &=& \left(Q^2-s_i+m_i^2\right)/\left(2\sqrt{Q^2}\right) \,,\\
   p_3^x &=&\sqrt{E_3^2-m_3^2} \,,\\
   p_1^x &=&\left(2 E_1E_3-s_2+m_1^2+m_3^2\right)/\left(2p_3^x\right) \,,\\
   p_2^x &=&\left(2 E_2E_3-s_1+m_2^2+m_3^2\right)/\left(2p_3^x\right) \,, \\
   p_1^y &=& \sqrt{E_1^2-\left(p_1^x\right)^2-m_1^2} \,,\\
   p_2^y &=& -\sqrt{E_2^2-\left(p_2^x\right)^2-m_2^2}=-p_1^y \,.
\eea
$s_3$ is defined analogously to $s_1$ and $s_2$ [i.e.,
  $s_3=(p_1+p_2)^2$] and may be expressed in terms of $s_1$, $s_2$ and
$Q^2$. Note that the angle between $\vec{p}_1$ and $\vec{p}_3$ is
fixed for a given choice of $s_1$, $s_2$ and $Q^2$.

The above definitions for the various hadron momentum vectors allow us
to determine simple expressions for $\widetilde{H}^{\mu\nu}$ in
$S$. We will not write out all 16 elements of the tensor. Rather, we
define new quantities $B_i$ ($i=1,\ldots, 4$) that are related to the
components of $\widetilde{J}^\mu$ as follows,
\bea
   B_1 &=& \widetilde{J}^1 = \left[F_1(p_1-p_3)^x+
     F_2(p_2-p_3)^x\right] \,, 
\label{eq:b1} \\
   B_2 &=& \widetilde{J}^2 = \left(F_1-F_2\right)p_1^y \,, \\
   B_3 &=& -i\widetilde{J}^3 = F_3 \sqrt{Q^2}p_1^yp_3^x\,, \\
   B_4 &=& \widetilde{J}^0 = \sqrt{Q^2}\left[
     F_4+\frac{f_H}{m_\tau}\eta_P\right] \, .
\label{eq:b4}
\eea
Then the components of $\widetilde{H}^{\mu\nu}$ consist of various
combinations $B_iB_j^*$ (in some cases multiplied by $\pm i$).
Inserting these expressions into Eq.~(\ref{eq:dgam}) and integrating
over $\alpha$ we find an expression very similar to that given in
Ref.~\cite{kilian},\footnote{Due to some ambiguities, it is difficult
  to tell if the expressions agree exactly.}
\bea
   && \frac{d\Gamma}{dQ^2\,ds_1 \,ds_2\,d\gamma \,d\cos\beta \,d\cos\theta} 
     =
     \frac{G_F^2\sin^2\!\theta_c}{512\left(2\pi\right)^6}
     \frac{\left(m_\tau^2-Q^2\right)^2}{m_\tau^3 Q^2} \nn \\
    &&~~~~~~~ \times\left\{
      \left[\frac{2}{3}K_1+K_2+
        \frac{1}{3}\overline{K}_1\left(3\cos^2\!\beta-1\right)\!/2\right]
	\left(\left|B_1\right|^2+\left|B_2\right|^2\right) 
	\rule{0pt}{15pt}\right. \nn \\
    &&~~~~~~~ ~~~~+\left[\frac{2}{3}K_1+K_2-
        \frac{2}{3}\overline{K}_1\left(3\cos^2\!\beta-1\right)\!/2\right]
	\left|B_3\right|^2 + K_2 \left|B_4\right|^2
	\nn \\
    &&~~~~~~~ ~~~~-\frac{1}{2}\overline{K}_1\sin^2\!\beta\cos 2\gamma
        \left(\left|B_1\right|^2-\left|B_2\right|^2\right)
	+ \overline{K}_1\sin^2\!\beta\sin 2\gamma\,
	    \mbox{Re}\!\left(B_1B_2^*\right)
	\nn \\
    &&~~~~~~~ ~~~~+2\overline{K}_3\sin\beta\sin\gamma\,
	    \mbox{Re}\!\left(B_1B_3^*\right)
	    +2\overline{K}_2\sin\beta\cos\gamma\,
	    \mbox{Re}\!\left(B_1B_4^*\right)
	\nn \\
    &&~~~~~~~ ~~~~+2\overline{K}_3\sin\beta\cos\gamma\,
	    \mbox{Re}\!\left(B_2B_3^*\right)
	    -2\overline{K}_2\sin\beta\sin\gamma\,
	    \mbox{Re}\!\left(B_2B_4^*\right)
	\nn \\
    &&~~~~~~~ ~~~~+2\overline{K}_3\cos\beta\,
	    \mbox{Im}\!\left(B_1B_2^*\right)
            +\overline{K}_1\sin 2\beta\cos\gamma\,
	    \mbox{Im}\!\left(B_1B_3^*\right)
	\nn \\
    &&~~~~~~~ ~~~\left. \rule{0pt}{15pt}
	-\overline{K}_1\sin 2\beta\sin\gamma\,
	    \mbox{Im}\!\left(B_2B_3^*\right)
            +2\overline{K}_2\cos\beta\,
	    \mbox{Im}\!\left(B_3B_4^*\right)\right\} \, ,
\label{eq:dgam2}
\eea
Note that, of the four parameters $B_i$ defined in
Eqs.~(\ref{eq:b1})-(\ref{eq:b4}), only $B_4$ contains a non-SM weak
phase. Thus, the only terms in Eq.~(\ref{eq:dgam2}) that can lead to
non-zero CP asymmetries are those containing one or more powers of
$B_4$. The parameters $K_i$ and $\overline{K}_i$ in the above
expression are defined as follows~\cite{KM1992},
\bea
   K_1 &=& 1-P\cos\theta -\left(m_\tau^2/Q^2\right)(1+P\cos\theta) \,,\\
   K_2 &=& \left(m_\tau^2/Q^2\right)(1+P\cos\theta) \,,\\
   K_3 &=& 1-P\cos\theta \,,\\
   \overline{K}_1 &=& K_1\left(3\cos^2\!\psi-1\right)\!/2
       -(3/2)K_4\sin 2\psi \,,\\
   \overline{K}_2 &=& K_2\cos\psi+K_4\sin\psi \,,\\
   \overline{K}_3 &=& K_3\cos\psi -K_4\sin\psi \,,\\
   K_4 &=& \sqrt{m_\tau^2/Q^2}\,P\sin\theta \,,
\eea
where the parameter $P$ denotes the polarization of the $\tau^-$,
$s_\tau^2=-P^2$. In the numerical work in Ref.~\cite{kilian}, a value
of $P$ was used that was relevant for LEP. In our numerical work we
will take $P=0$, which is appropriate for lower-energy
experiments~\cite{KM1992}. Note that $\psi$ is a function of
$\cos\theta$ and $Q^2$. If the $\tau$'s are pair-produced at a
symmetric collider,
\bea
   \cos\theta &=& \frac{\left(2xm_\tau^2-m_\tau^2-Q^2\right)}
    {\left(m_\tau^2-Q^2\right)\sqrt{1-4m_\tau^2/s}} \,, 
    \label{eq:costheta}\\
   \cos\psi &=& \frac{x\left(m_\tau^2+Q^2\right)-2Q^2}
    {\left(m_\tau^2-Q^2\right)\sqrt{x^2-4Q^2/s}}\,,
    \label{eq:cospsi}
\eea
where $x=2E_h/\sqrt{s}$ and $s=4E_{\textrm{\scriptsize{beam}}}^2$,
with $E_h$ being the hadron energy in the lab~\cite{KM1992} (see also
Ref.~\cite{KW1984}). Thus, given $s$ (we take
$s=\left(10.58~\mbox{GeV}\right)^2$ below), $Q^2$ and $\cos\theta$,
one can solve for $x$ and substitute this expression into the
expression for $\cos\psi$. Finally, note that if the direction of the
$\tau^-$ could be determined, it would not be necessary to integrate
over $\alpha$.  In this case it might be possible to extract other
useful information for the construction of CP asymmetries. We do not
consider this possibility in this work.

The differential width in Eq.~(\ref{eq:dgam2}) may now be integrated
to compute the partial width for $\tau^-\to K^-\pi^-\pi^+\nu_\tau$.
Comparison with the analogous quantity for the $\tau^+$ decay yields
the regular rate asymmetry.  One can also integrate over the angular
variables in an asymmetric manner in such a way that certain cross
terms are selected from Eq.~(\ref{eq:dgam2}).  These ``weighted
differential widths'' can then be compared to the analogous
expressions for the $\tau^+$ decay to yield CP-odd quantities.  We
consider two types of asymmetries formed in this manner -- modified
rate asymmetries [whose dependence on the strong and weak phases is
  given in Eq.~(\ref{rateangles})] and a triple product asymmetry [see
  Eq.~(\ref{TPangles})].

\subsection{Weighted differential widths}

The authors of Ref.~\cite{kilian} derived an expression for the
differential width that is very similar to Eq.~(\ref{eq:dgam2}).
Since they assumed that $f_H=0$ for these decays, they only considered
LR effects.  We consider the complementary point of view. Assuming
that $f_H$ could be non-zero and noting that there are strong
constraints on LR mixing, we consider only effects due to the exchange
of a charged scalar.  The analysis in Ref.~\cite{kilian} focused
exclusively on triple products in the differential width. In our
notation, these TP's correspond to the cross-terms containing the
factors Im$(B_iB_j^*)$. Recall that CP asymmetries formed from triple
products do not require the presence of strong phases [see
  Eq.~(\ref{TPangles})].

In this work, we reconsider CP asymmetries formed from triple products
and also consider CP asymmetries that can be formed from
T-even\footnote{``T-even'' here refers to the naive time-reversal
  operation.} cross-terms in the differential width.  Both types of
terms may be isolated by employing suitable weighting functions when
performing the angular integrations.

We begin by defining various regions in terms of $\gamma$ and $\beta$,
as in Ref.~\cite{kilian},
\bea
\begin{array}{ll}
  \mbox{I}:~0\leq \gamma< \pi/2, & \mbox{II}:~\pi/2\leq \gamma<\pi, \\
  \mbox{III}:~\pi\leq \gamma< 3\pi/2, &\mbox{IV}:~3\pi/2\leq \gamma<2\pi;\\
  & \\
  \mbox{A}: ~0\leq \beta< \pi/2, & \mbox{B}:~\pi/2\leq \beta<\pi; \\
\end{array}
\label{eq:regions}
\eea
As noted above, in order for a particular term in the differential
width [Eq.~(\ref{eq:dgam2})] to contribute to a non-zero CP asymmetry,
it must contain one or more powers of $B_4$. This is because $B_4$
contains the possible CP-violating phase coming from NP. Inspection of
Eq.~(\ref{eq:dgam2}) leads one to the conclusion that there are four
terms of interest. One is proportional to $\left|B_4\right|^2$. As we
shall see below, this term arises in the regular rate asymmetry. The
remaining three terms are proportional to the angular functions
$\sin\beta\cos\gamma$, $\sin\beta\sin\gamma$ and $\cos\beta$. These
three terms can be isolated by using appropriate weighting functions,
as indicated in Table~\ref{tab:angular}. Thus, for example, to isolate
the term in Eq.~(\ref{eq:dgam2}) proportional to $\sin \beta
\sin\gamma$, the differential width is multiplied by
$g_1(\gamma,\beta)$ (which is $+1$ in regions IA, IIA, IB and IIB and
$-1$ in the other regions) and the angular integration is carried
out. This eliminates all other terms since the weighting functions are
such that\footnote{The weighting functions are also orthogonal to $1$,
  $\left(3\cos^2\beta -1\right)/2$, $\sin^2\beta\cos 2\gamma$, etc.,
  so that only the intended cross-terms are isolated. Also note that,
  experimentally, a more statistically significant weighting procedure
  might be to weight the differential width by the various functional
  forms $f_i$ themselves. See Ref.~\cite{wkn2} and also the moment
  analysis discussion in Ref.~\cite{KM1992}.}
\bea
   \int\!\!\int f_i(\gamma,\beta) g_j(\gamma,\beta) 
        \sin\beta \,d\gamma \,d\beta=2\pi\,\delta_{ij} \,~~~~~ (i=1,2,3)\,.
\eea
\begin{table*}
\caption{Angular weighting factors. The regions I-IV, A and B are
  defined in Eq.~(\ref{eq:regions}) in the text. The second column
  gives the angular functions of interest, $f_i(\gamma,\beta)$. The
  third column gives the weighting function $g_i(\gamma,\beta)$ that
  can be used to isolate $f_i(\gamma,\beta)$. The functions
  $g_i(\gamma,\beta)$ are simply $\pm 1$ depending on which region
  $\gamma$ and $\beta$ fall in.}
\label{tab:angular}
\begin{center}
\begin{tabular}{c|cc} \hline\hline 
  \rule{0pt}{15pt} 
  $i$ & $f_i(\gamma,\beta)$ & $g_i(\gamma,\beta)$ \\ \hline
  \rule{0pt}{15pt} 
  1 & $\sin\beta\sin\gamma$ & 
    $\mbox{I}+\mbox{II}-
       \mbox{III}-\mbox{IV};\mbox{A}+\mbox{B}$ 
    \\
  \rule{0pt}{0pt} 
  2 & $\sin\beta\cos\gamma$ & 
    $\mbox{I}-\mbox{II}-
       \mbox{III}+\mbox{IV};\mbox{A}+\mbox{B}$ 
    \\ \hline
  \rule{0pt}{0pt} 
  3 & $\cos\beta$ & 
    $\mbox{I}+\mbox{II}+
       \mbox{III}+\mbox{IV};\mbox{A}-\mbox{B}$ 
    \\ \hline \hline
\end{tabular}
\end{center}
\end{table*}

Using the weighting functions $g_i(\gamma,\beta)$ in
Table~\ref{tab:angular}, we define weighted differential widths as
follows,
\bea
   \frac{d\Gamma_i}{dQ^2\,ds_1 \,ds_2}\equiv
    \int \frac{d\Gamma}{dQ^2\,ds_1 \,ds_2\,d\gamma \,d\cos\beta \,d\cos\theta}
    \, g_i(\gamma,\beta) \sin\beta \, d\beta \, d\gamma\, d\cos\theta \,.
    \label{eq:weighteddifflwidths}
\eea
The results for the three weighting functions are as follows,
\bea
   \frac{d\Gamma_1}{dQ^2\,ds_1 \,ds_2} &=& A(Q^2)
     \left[\langle \overline{K}_3\rangle \mbox{Re}\!\left(B_1B_3^*\right)
       -\langle \overline{K}_2\rangle \mbox{Re}\!\left(B_2B_4^*\right)
         \right]\,,
\label{eq:dgam1}\\
   \frac{d\Gamma_2}{dQ^2\,ds_1 \,ds_2} &=& A(Q^2)
     \left[\langle \overline{K}_3\rangle \mbox{Re}\!\left(B_2B_3^*\right)
       +\langle \overline{K}_2\rangle \mbox{Re}\!\left(B_1B_4^*\right)
         \right]\,,
         \label{eq:dgam2a}\\
   \frac{d\Gamma_3}{dQ^2\,ds_1 \,ds_2} &=& A(Q^2)
     \left[\langle \overline{K}_3\rangle \mbox{Im}\!\left(B_1B_2^*\right)
       +\langle \overline{K}_2\rangle \mbox{Im}\!\left(B_3B_4^*\right)
         \right]\,,
\label{eq:dgam3}	
\eea
where
\bea
   A(Q^2) &=& \frac{G_F^2\sin^2\!\theta_c}{128\left(2\pi\right)^5}
     \frac{\left(m_\tau^2-Q^2\right)^2}{m_\tau^3 Q^2} \,,\\
   \langle \overline{K}_i\rangle &\equiv &
      \frac{1}{2}\int_0^\pi \overline{K}_i \sin\theta \, d\theta \,.
\eea

The three weighted differential widths defined in
Eqs.~(\ref{eq:dgam1})-(\ref{eq:dgam3}) can now be compared to their
CP-conjugates in order to construct CP asymmetries.  Recalling that
the Higgs contribution resides in $B_4$ [see Eq.~(\ref{eq:b4})] and
noting that each of the three expressions above contains a term linear
in $B_4$, we see that each of the resulting CP asymmetries has the
possibility of being non-zero.

In the following sections we construct the CP asymmetries and then
study them numerically to see if they might provide useful probes of
non-SM CP violation.

\section{CP-odd observables}
\label{sec:cpodd}

Before analyzing the various CP asymmetries, let us consider the
coefficients $B_i$ defined in Eqs.~(\ref{eq:b1})-(\ref{eq:b4}) a bit
more carefully.  As noted above, the sole non-SM weak phase resides in
$B_4$.  The form factors $F_i$ and $f_H$ are potential sources of
strong phases.  We may thus parameterize the four coefficients as
follows,
\bea
   B_1 & = & \left|B_1\right|e^{i\delta_1} \,,
\label{eq:b1a} \\
   B_2 & = & \left|B_2\right|e^{i\delta_2} \,,
\label{eq:b2a} \\
   B_3 & = & \left|B_3\right|e^{i\delta_3} \,,
\label{eq:b3a} \\
   B_4 & = & \left|B_4^{(1)}\right|e^{i\delta_4}
    +\left|B_4^{(2)}\right|e^{i\delta_H+i\phi_H}\,,
\label{eq:b4a}
\eea
where
\bea
   B_4^{(1)} = \sqrt{Q^2}F_4\,,~~~
   B_4^{(2)} = \sqrt{Q^2}\frac{f_H}{m_\tau}\eta_P
     \, ,
     \label{eq:b42}
\eea
and where $\delta_i$ and $\phi_H$ represent strong and weak phases,
respectively. An explicit expression for the weak phase $\phi_H$ is as
follows,
\bea
   \phi_H &=& \arg\left(\eta_P\right)\,.
\eea
This phase could in principle be of order unity.

As was the case in Ref.~\cite{DeltaS=0}, we can consider three types
of CP asymmetries.  The first is the regular rate asymmetry.  This
asymmetry is likely to be small in $\tau\to K\pi\pi\nu_\tau$ and is
therefore unlikely to be measureable in the near future.  The second
and third types of asymmetries are the modified rate asymmetry and the
triple-product asymmetry.  We consider two different modified rate
asymmetries, and one triple-product asymmetry.  The triple-product
asymmetry is similar to one considered for the decay $\tau\to K\pi
K\nu_\tau$ in Ref.~\cite{kilian}.  The modified rate asymmetries, to
our knowledge, are new relative to this decay mode.  Both types of
asymmetries are constructed by first performing an asymmetrical
integration over the kinematical angles $\beta$ and $\gamma$, as noted
in Eq.~(\ref{eq:weighteddifflwidths}) and Table~\ref{tab:angular}.
Since the procedures for extracting these two types of asymmetries are
similar, we consider them together in the following.

\subsection{Rate asymmetry}

Let us first consider the regular rate asymmetry. In this case the
angular integrations are performed symmetrically [$g(\gamma,\beta)=1$]
and the width for the process is compared to that for the
anti-process.  The differential width for the $\tau^-$ decay in this
case is given by
\bea
   \frac{d\Gamma}{dQ^2\,ds_1 \,ds_2}= A(Q^2)
    \left[\left(\frac{2}{3}+\frac{1}{3}\frac{m_\tau^2}{Q^2}\right)
      \left(\left|B_1\right|^2+\left|B_2\right|^2+\left|B_3\right|^2\right)
      +\frac{m_\tau^2}{Q^2}\left|B_4\right|^2\right] \,.
\label{eq:dgam_no_asym}
\eea
The width for the $\tau^+$ process will have the same strong phases,
but the weak phases will have their signs reversed. It is immediately
evident from Eq.~(\ref{eq:dgam_no_asym}) and
Eqs.~(\ref{eq:b1a})-(\ref{eq:b3a}) that the coefficients $B_1$, $B_2$
and $B_3$ will not give any contribution to the rate asymmetry, since
they do not contain weak phases. Thus, the rate asymmetry is
proportional to\footnote{This expression is part of an integral over
  phase space. Note that one or both of the strong phases could depend
  on $Q^2$, $s_1$ and $s_2$.}
\bea
   \left|B_4\right|^2- \left|\overline{B}_4\right|^2
    = 4 \left|B_4^{(1)}\right|\left|B_4^{(2)}\right|
     \sin\left(\delta_4-\delta_H\right)
      \sin\left(\phi_H\right) \,.
\eea
This expression is proportional to $\left|F_4 f_H \eta_P\right|$. The
SM scalar form factor $F_4$ is generally thought to be small.  If the
NP factor $\eta_P$ is also small, then the regular rate asymmetry is
doubly suppressed.  Given the expected smallness of the regular rate
asymmetry, we do not consider it further here. As we shall see,
however, other CP asymmetries can be constructed that depend on
$F_if_H\eta_P$, with $i=1, 2, 3$.  Such asymmetries may
be more promising in terms of their NP reach.

\subsection{Modified and triple-product CP asymmetries}

We define CP asymmetries corresponding to the weighted differential
widths [Eqs. (\ref{eq:dgam1})-(\ref{eq:dgam3})] as follows,
\bea
   A_{CP}^{(i)} = \frac{1}{\Gamma +\overline{\Gamma}}
     \int \left(\frac{d\Gamma_i}{dQ^2\,ds_1 \,ds_2} 
      -\frac{d\overline{\Gamma}_i}{dQ^2\,ds_1 \,ds_2}\right)
      dQ^2\,ds_1 \,ds_2 \,.
\label{eq:ACP} 
\eea
The quantities with the bars correspond to the decay $\tau^+\to K^+
\pi^+\pi^-{\bar\nu}_\tau$ and are obtained from those without the bars
by changing the signs of all weak phases while leaving strong phases
unchanged.\footnote{Note that we {\em subtract} the width for the
  anti-process from that for the process, both for the modified rate
  asymmetries and for the triple-product asymmetry. The authors of
  Ref.~\cite{kilian} consider only triple-product asymmetries. Their
  expressions for the anti-process contain an extra over-all sign;
  thus they add the widths for the process and anti-process to obtain
  CP asymmetries. This is a notational difference.  Both approaches
  lead (correctly) to a TP CP asymmetry that is of the form of
  Eq.~(\ref{TPangles}).}  $A_{CP}^{(1)}$ and $A_{CP}^{(2)}$ descend
from the terms containing Re$(B_2B_4^*)$ and Re$(B_1B_4^*)$ in
Eqs.~(\ref{eq:dgam1}) and (\ref{eq:dgam2a}), respectively.  These are
both modified rate asymmetries.  The third asymmetry, $A_{CP}^{(3)}$,
descends from the term containing Im$(B_3B_4^*)$ in
Eq.~(\ref{eq:dgam3}).  This a triple-product asymmetry.  $\Gamma$ and
$\overline{\Gamma}$ in Eq.~(\ref{eq:ACP}) represent the partial widths
for $\tau^-\to K^-\pi^-\pi^+\nu_\tau$ and $\tau^+\to
K^+\pi^+\pi^-\overline{\nu}_\tau$, respectively.  In our numerical
work below we make the approximation that $\Gamma\simeq
\overline{\Gamma}$, so that $\Gamma +\overline{\Gamma}\simeq 2\Gamma.$
The experimental value for $\Gamma$ is used.

\subsubsection{Modified rate asymmetries ($i=1,2$)}

The modified rate asymmetries, $A_{CP}^{(1)}$ and $A_{CP}^{(2)}$,
require a strong phase in order to be non-zero. These asymmetries are
analogous to the ``polarization-dependent asymmetry'' defined in
Ref.~\cite{DeltaS=0}.  In order to obtain numerical estimates for
these asymmetries, let us make the following simplifying
assumptions. First of all, we will assume that $f_H$ has no $Q^2$,
$s_1$ or $s_2$ dependence. We will also assume that $f_H$ has no
strong phase associated with it (it will be taken to be real and
positive). Under these assumptions, these two asymmetries are given by
\bea
   A_{CP}^{(1)} & \simeq & -\frac{m_\tau}{\Gamma +\overline{\Gamma}}
    \left[
     \int \frac{A(Q^2)}{\sqrt{Q^2}}\cos\psi \, p_1^y\,
      \mbox{Im}\!\left(F_1-F_2\right)dQ^2 ds_1ds_2\,d\!\cos\theta
     \right] \nonumber \\
    && ~~~~\times f_H \,\mbox{Im}\!\left(\eta_P\right) \,,
    \label{eq:Acp1}\\
    && \nonumber\\
   A_{CP}^{(2)} & \simeq & \frac{m_\tau}{\Gamma +\overline{\Gamma}}
    \left[
     \int \frac{A(Q^2)}{\sqrt{Q^2}}\cos\psi \,
      \mbox{Im}\!\left[F_1\left(p_1-p_3\right)^x+
	F_2\left(p_2-p_3\right)^x\right]dQ^2 ds_1ds_2\,d\!\cos\theta
     \right] \nonumber \\
    && ~~~~\times f_H \,\mbox{Im}\!\left(\eta_P\right) \,,
    \label{eq:Acp2}
\eea
in which we have taken the $\tau$'s to be unpolarized ($P=0$). Recall
that $\psi$ depends on $\theta$ through Eqs.~(\ref{eq:costheta}) and
(\ref{eq:cospsi}).

As noted above, $A_{CP}^{(1)}$ and $A_{CP}^{(2)}$ both have the
generic form $\sin\phi\sin\delta$, since
$\mbox{Im}\!\left(\eta_P\right) \propto\sin\phi_H$ and
$\mbox{Im}\!\left(F_1-F_2\right)$ and
$\mbox{Im}\!\left[F_1\left(p_1-p_3\right)^x+
  F_2\left(p_2-p_3\right)^x\right]$ are both proportional to
$\sin\delta$, with $\delta$ being a strong phase. In
Sec.~\ref{sec:numerical} we will examine the sensitivity of these
asymmetries in a particular model for the form factors.

\subsubsection{Triple-product asymmetry ($i=3$)}

The third CP asymmetry, $A_{CP}^{(3)}$, is a triple-product asymmetry
and is similar in some respects to the asymmetries constructed for
$\tau\to K \pi\pi \nu_\tau$ in Ref.~\cite{kilian}.  Recall, however,
that in that case the authors assumed that the NP effects were due to
a new right-handed gauge boson. To obtain a numerical estimate for
$A_{CP}^{(3)}$, we make the same simplifying assumptions as above;
i.e., we assume that $f_H$ is real and positive (no strong phase) and
that it has no $Q^2$, $s_1$ or $s_2$ dependence.  Under these
assumptions,
\bea
   A_{CP}^{(3)} & \simeq & -\frac{m_\tau}{\Gamma +\overline{\Gamma}}
    \left[
     \int A(Q^2)\cos\psi \, p_1^y \, p_3^x\,
      \mbox{Re}\!\left(F_3\right)dQ^2 ds_1ds_2\,d\!\cos\theta
     \right] \nonumber\\
    && ~~~~\times f_H \,\mbox{Im}\!\left(\eta_P\right) \,.
    \label{eq:Acp3}
\eea
Like the modified rate asymmetries considered above, the
triple-product asymmetry $A_{CP}^{(3)}$ is proportional to
$\mbox{Im}\!\left(\eta_P\right)$. In contrast to $A_{CP}^{(1)}$ and
$A_{CP}^{(2)}$, however, this asymmetry does not require a strong
phase, since $\mbox{Re}\!\left(F_3\right)\propto\cos\delta$ (where
$\delta$ represents a strong phase). Having said this, there is a
potential drawback with $A_{CP}^{(3)}$ in that it depends on the
sub-dominant Wess-Zumino form factor $F_3$, whereas $A_{CP}^{(1)}$ and
$A_{CP}^{(2)}$ depend on combinations of the dominant form factors
$F_1$ and $F_2$.  In the next section we perform a numerical study to
examine these various factors quantitatively.

\section{Numerical Results}
\label{sec:numresults}

The modified and triple-product asymmetries defined above all have the
form
\bea A_{CP}^{(i)} = a_{CP}^{(i)} f_H
   \mbox{Im}\!\left(\eta_P\right)\,,~~~~i=\mbox{1--3}\,, 
\label{eq:asym_form}
\eea
where the $a_{CP}^{(i)}$ are constants determined by integrating over
$\cos\theta$, $s_1$, $s_2$ and $Q^2$. In this section we assume
particular functional forms for the form factors and use these to
estimate the $a_{CP}^{(i)}$.  It turns out that there are significant
cancellations that occur as one performs the integrations over phase
space.  To help illustrate this cancellation, we define four
differential quantities as follows,
\beq \frac{da_{CP}^{(i)}}{dX}\,, 
\label{eq:diff_asym}
\eeq
with $X$ given by $M_{K\pi\pi} = \sqrt{Q^2}$, $M_{\pi\pi} =
\sqrt{s_1}$, $M_{K\pi} = \sqrt{s_2}$ and $\cos\theta$.  Given the
cancellations that occur upon integration, experimentalists may wish
to study differential CP asymmetries in addition to, or in place of,
the integrated asymmetries.

\subsection{Model for the form factors}
\label{sec:ff}

There have been several models for the form factors $F_1$-$F_3$ over
the past number of years.  One model, which simply took the
intermediate $K_1$ to be the $K_1(1400)$, may be found in
Ref.~\cite{dmsw} (see also Ref.~\cite{jadach}).  A subsequent analysis
by Finkemeier and Mirkes~\cite{finkemeiermirkes1} took into account
both the $K_1(1400)$ and the $K_1(1270)$ resonances and also
incorporated other $K^*$ resonances ($K^{*\prime}$ and
$K^{*\prime\prime}$; see also Refs.~\cite{portoles,jamin}).  Finally,
an experimental analysis of the form factors was performed by the CLEO
collaboration in Ref.~\cite{CLEO_K1}.

The various models that have been proposed make different assumptions
regarding the anomalous Wess-Zumino term, $F_3$.  The authors of
Ref.~\cite{dmsw} found that the $F_3$ term contributes approximately
$1\%$ to the overall width for $\tau^-\to K^-\pi^-\pi^+\nu_\tau$.
The parameterization in Ref.~\cite{finkemeiermirkes1} led to an
anomalous contribution of order $10\%$.  The CLEO collaboration noted
that the contribution would be of order $5.5\%$ based on a particular
model (found in Ref.~\cite{jadach}).  Since the contribution was
expected to be small, they set $F_3$ to zero in their analysis and
focused on determining the resonance structures of $F_1$ and $F_2$.
The uncertainty resulting from the neglect of $F_3$ was incorporated
into their systematic error~\cite{CLEO_K1}.

We model our numerical work after the CLEO analysis, with the main
exception being that we allow $F_3$ to be non-zero.  Guided by
Ref.~\cite{CLEO_K1} for $F_1$ and $F_2$ and by Ref.~\cite{dmsw} for
$F_3$, we write the form factors in terms of various Breit-Wigner
functions as follows,
\bea
   F_1(s_1,s_2,Q^2) & = & - \frac{2N}{3 F_\pi}\left[C\cdot BW_{1270}(Q^2)
    + D\cdot BW_{1400}(Q^2)\right]BW_{K^*}(s_2) 
\label{eq:f1}\,, \\
   F_2(s_1,s_2,Q^2) & = & - \frac{N}{\sqrt{3}F_\pi}\left[A\cdot BW_{1270} (Q^2)
    + B\cdot BW_{1400}(Q^2)\right] T_\rho^{(1)}(s_1) 
\label{eq:f2}\,, \\
   F_3(s_1,s_2,Q^2) & = & \frac{N_3}{2\sqrt{2}\pi^2 F_\pi^3}
    BW_{K^*}(Q^2)\times \left[\frac{T_\rho^{(1)}(s_1)+\alpha BW_{K^*}(s_2)}
     {1+\alpha}\right] \,,
\label{eq:f3}
\eea
with $\alpha = -0.2$ and $F_\pi = 93.3$~MeV.  Also, we set $F_4$ to
zero and only take $f_H$ into account when computing the numerators of
the asymmetry expressions.  The constants $N$, $N_3$ and $A$-$D$ will
be discussed further below.  The normalized Breit-Wigner propagators
for the $K_1(1270)$ and the $K_1(1400)$ are assumed to be given
by~\cite{CLEO_K1},
\bea
   BW_{K_1} (Q^2) = \frac{-m_{K_1}^2+im_{K_1}\Gamma_{K_1}}
    {Q^2-m_{K_1}^2+im_{K_1}\Gamma_{K_1}} \,,
\eea
with $m_{K_1}$ and $\Gamma_{K_1}$ being the mass and width for the
appropriate $K_1$ state.  As noted in the CLEO analysis, a fit to the
$\tau\to K\pi\pi\nu_\tau$ data indicates that the effective
$K_1(1270)$ and $K_1(1400)$ widths are larger in this decay than the
respective values reported by the Particle Data Group (see also
Refs.~\cite{finkemeierkuhnmirkes1997,kuhnetal1997}).  Following CLEO,
we take the following values for our numerical
analysis~\cite{CLEO_K1}:
\bea
   m_{1270\,(1400)} = 1.254\,(1.463)~\mbox{GeV},
   ~~~\Gamma_{1270\,(1400)} = 0.26\,(0.30)~\mbox{GeV}.
\eea
The Breit-Wigner propagators for the $K^*$ and $\rho$ are taken to
have energy-dependent widths (see, for example,
Refs.~\cite{CLEO_K1,dmsw}),
\bea
   BW_R (s) = \frac{-m_R^2}{s-m_R^2+i\sqrt{s}\Gamma_R(s)}\,.
\eea
with
\bea
   \Gamma_R(s) = \Gamma_R \frac{m_R^2}{s}\left(\frac{p}{p_R}\right)^3\,,
   \label{eq:gammas}
\eea
where 
\bea
   p &=& \frac{1}{2\sqrt{s}}\sqrt{[s-(m_1+m_2)^2][s-(m_1-m_2)^2]} \,,\\
   p_R &=& \frac{1}{2m_R}
    \sqrt{[m_R^2-(m_1+m_2)^2][m_R^2-(m_1-m_2)^2]} \,.
\eea
When using the above expressions it is assumed that the resonance $R$
decays to two particles with masses $m_1$ and $m_2$.
[Equation~(\ref{eq:gammas}) also assumes that $\sqrt{s}\geq m_1+m_2$
  -- otherwise $\Gamma_R(s)$ should be set to zero.  This condition is
  satisfied in all regions of phase space for the decay chains that we
  consider.]  For the $K^*$, a single resonance (with an
energy-dependent width) is assumed; we take $m_{K^*}=0.892$~GeV and
$\Gamma_{K^*}=0.050$~GeV.\footnote{Note that the intermediate $K^*$
  represents a $K^{*0}$ in the expression for the form factor $F_1$,
  while both $K^{*0}$ and $K^{*-}$ appear in $F_3$.  For simplicity we
  use the same mass and width for both the charged and neutral
  versions of this particle.}  The expression for the $\rho$
incorporates two different resonances (the $\rho$ and the
$\rho^\prime$),
\bea
   T_\rho^{(1)} (s_1) = \frac{BW_\rho(s_1)+\beta BW_{\rho^\prime}(s_1)}
    {1+\beta} \,,
\eea
with $\beta=-0.145$, $m_\rho=0.773$~GeV, $m_{\rho^\prime} =
1.370$~GeV, $\Gamma_\rho=0.145$~GeV and $\Gamma_{\rho^\prime} =
0.510$~GeV~\cite{finkemeiermirkes1,kuhnsantamaria}.

Let us now consider the constants $N$, $N_3$ and $A$-$D$ in
Eqs.~(\ref{eq:f1})-(\ref{eq:f3}).  The CLEO collaboration effectively
set $N_3=0$ in their analysis and then determined
$A$-$D$~\cite{CLEO_K1}.  The overall normalization $N$ was not stated.
We choose values that are similar to those reported in Table~I of
Ref.~\cite{CLEO_K1},
\bea
   A &=& 0.944\;, \nn\\
   B &=& 0\;,\nn\\
   C &=& A\times\sqrt{\frac{16}{42}}\times\sqrt{\frac{6917}{61636}}\simeq
   0.195\;,\nn\\
   D &=& \sqrt{1-A^2-C^2}\simeq 0.266\; .
\eea
An apparent typo in Eq.~(2) of Ref.~\cite{CLEO_K1} renders
the relative signs of the constants a bit uncertain.  The
signs we have chosen for $A$-$D$ are consistent with the
signs used in Ref.~\cite{finkemeiermirkes1}.  Our parameter
choice gives results for the differential width (see
Fig.~\ref{fig:dgam} below) that are visually similar to the
results obtained in Ref.~\cite{CLEO_K1}, although the
agreement between our numerical results and those of CLEO is
not perfect.  Since we wish, in part, to study effects due to
the inclusion of the $F_3$ term, we retain a non-zero value
for $N_3$.  As was noted above, there have been various
estimates regarding the $F_3$ contribution to the width, with
estimates varying from $1\%$ to $10\%$ in papers that we have
noted.  For the purpose of our numerical study, we fix $N$
and $N_3$ such that the $F_3$ term contributes $5\%$ to the
$\tau\to K\pi\pi\nu_\tau$ width, with the $F_1$ and $F_2$
terms contributing the remaining $95\%$.  Taking ${\cal
B}(\tau\to K\pi\pi\nu_\tau)=0.00273$~\cite{BABAR_BR} and
$c\tau = 87.11\times 10^{-6}$~m~\cite{pdg}, we find that
$N\simeq 1.4088$ and $N_3\simeq 1.4696$.

%
\begin{figure}[t]
\begin{center}
\resizebox{5in}{!}{\includegraphics*{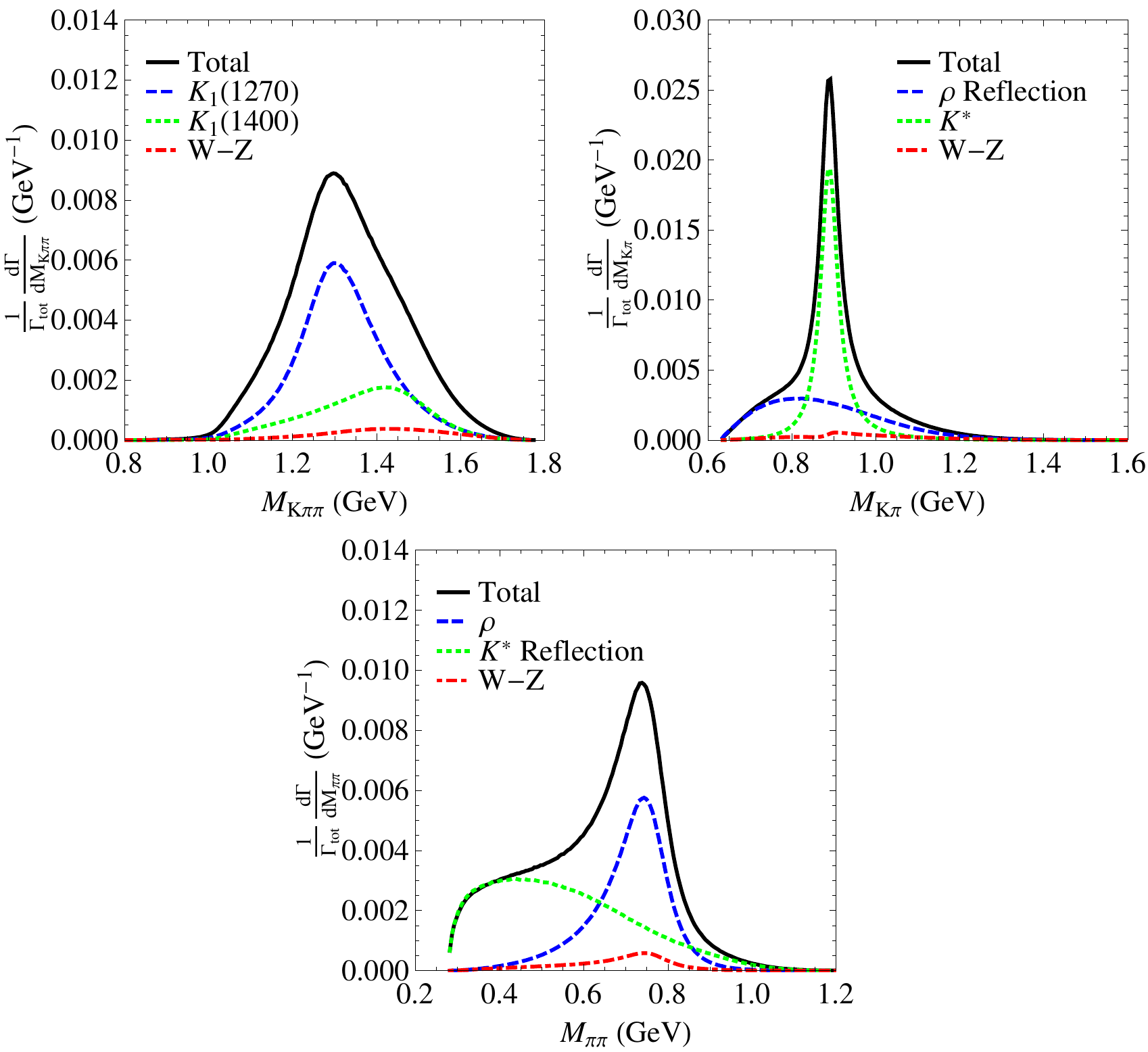}}
\caption{Plots of the differential widths $d\Gamma/dM$, including the
  contributions from the various decay chains.  The $\rho$, $K^*$,
  $K_1(1270)$ and $K_1(1400)$ curves include contributions from only
  the $F_1$ and $F_2$ terms.  The ``W-Z'' curves represent the
  contribution from the anomalous Wess-Zumino form factor, $F_3$.}
\label{fig:dgam}
\end{center}
\end{figure}
%

\subsection{Numerical estimates of the CP asymmetries}
\label{sec:numerical}

\begin{table*}
\caption{ Calculated values for $a_{CP}^{(i)}$,
  $a_{CP\textrm{\scriptsize{,mod}}}^{(i)}$, and
  $a_{CP\textrm{\scriptsize{,max}}}^{(i)}$.
  $a_{CP\textrm{\scriptsize{,mod}}}^{(i)}$ is computed by making the
  replacement $\cos\psi\to\left|\cos\psi\right|$ in
  Eqs.~(\ref{eq:Acp1})-(\ref{eq:Acp3}).  This procedure helps to
  eliminate some of the cancellations that occur upon integration.
  $a_{CP\textrm{\scriptsize{,max}}}^{(i)}$ is determined by taking the
  absolute values of the integrands in
  Eqs.~(\ref{eq:Acp1})-(\ref{eq:Acp3}).  }
\label{tab:asymmetries}
\begin{center}
\begin{tabular}{c|c|c|c|cc} \hline\hline 
  \rule{0pt}{15pt} 
  $i$ & $a_{CP}^{(i)}$ & $a_{CP\textrm{\scriptsize{,mod}}}^{(i)}$ 
    & $a_{CP\textrm{\scriptsize{,max}}}^{(i)}$ & CP asymmetry type \\ \hline
  \rule{0pt}{15pt} 
  1 & $-2.2 \times 10^{-5}$ &  $-5.2 \times 10^{-5}$ & $9.8 \times
  10^{-4}$ 
    & Modified rate asymmetry 
    \\
  \rule{0pt}{0pt} 
  2 & $7.0 \times 10^{-4}$ & $1.0 \times 10^{-3}$ & $2.9 \times
  10^{-3}$ 
    & Modified rate asymmetry
    \\ 
  \rule{0pt}{0pt} 
  3 & $2.5 \times 10^{-4}$ & $6.2 \times 10^{-4}$ & $8.3 \times
  10^{-4}$ 
    & Triple-product asymmetry
    \\ \hline \hline
\end{tabular}
\end{center}
\end{table*}

%
\begin{figure}[t]
\begin{center}
    \resizebox{5in}{!}{\includegraphics*{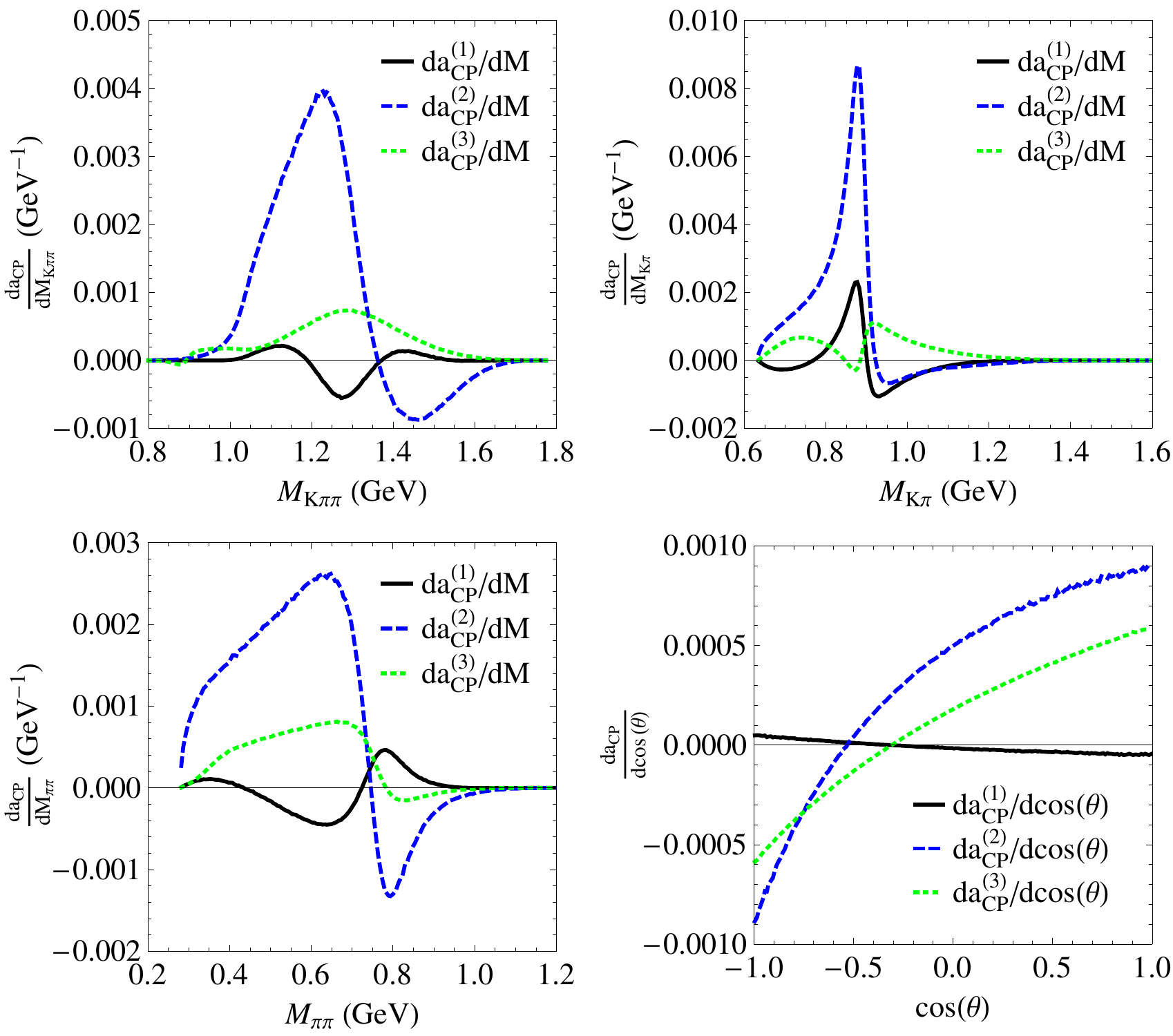}}
\caption{Differential asymmetries showing each asymmetry's dependence
  on the respective integration variables.}
\label{fig:dacp}
\end{center}
\end{figure}
%

Using the constants for $N$, $N_3$, and $A$-$D$ noted above, we
integrate Eq.~(\ref{eq:dgam2}) over phase space to obtain
$d\Gamma/dM_{K\pi\pi}$, $d\Gamma/dM_{K\pi}$ and $d\Gamma/dM_{\pi\pi}$.
The results (normalized to $\Gamma_{\textrm{\scriptsize{tot}}}$) are
displayed in Fig.~\ref{fig:dgam}.  The plots are similar to those in
Fig.~2 of Ref.~\cite{CLEO_K1}, although the agreement is not perfect.
We also include a Wess-Zumino contribution (not included in the CLEO
plots).

Having chosen the various coefficients, we can also perform the
integrations in Eqs.~(\ref{eq:Acp1})-(\ref{eq:Acp3}) to obtain the
numerical coefficients $a_{CP}^{(i)}$.  Numerical values for these
coefficients are listed in the second column of
Table~\ref{tab:asymmetries}.  Recall that the actual CP asymmetries
are obtained by multiplying the $a_{CP}^{(i)}$ by
$f_H$Im$\left(\eta_P\right)$ [see Eq.~(\ref{eq:asym_form})].

Figure~\ref{fig:dacp} shows plots of the differential asymmetries
$da_{CP}^{(i)}/dX$, with $X=M_{K\pi\pi}$, $M_{\pi\pi}$, $M_{K\pi}$ and
$\cos\theta$.  In each case, integration over $X$ yields the
corresponding coefficient $a_{CP}^{(i)}$.  As is apparent from the
figure, each of the asymmetry coefficients undergoes considerable
cancellation upon integration.  Given these cancellations,
experimentalists may find it advantageous to perform fits to the
differential CP asymmetries instead of simply measuring the integrated
asymmetries.  Alternatively, it may be possible to achieve larger
integrated asymmetries by employing extra weighting functions in the
integration over one or more of the integration variables.  As an
example, we have recomputed the asymmetries with the change
$\cos\psi\to \left| \cos\psi \right|$ in
Eqs.~(\ref{eq:Acp1})-(\ref{eq:Acp3}) (as noted above, $\cos\psi$
should be an experimental observable).  The third column of
Table~\ref{tab:asymmetries} shows the resulting values for the
asymmetry coefficients.  As can be seen, this modification leads to
modest increases in the sizes of the coefficients.  Other weighting
functions could also be considered.  If a weighting function is chosen
such that it takes on only the values $\pm 1$ over the integration
range, the largest possible asymmetry coefficients would be obtained
by simply taking the absolute value of the integrand.  We have
computed the asymmetry coefficients under this assumption as well.
The results may be found in the fourth column of
Table~\ref{tab:asymmetries}.  The values in this column represent the
maximum values obtainable for the magnitudes of the asymmetry
coefficients, given the choices we have made for the form factors.
Comparison of the second and fourth columns in the table illustrates
the level of cancellation that the integrated asymmetry coefficients
have each undergone.  A considerable gain in the magnitude of each
asymmetry is possible if an appropriate weighting function is adopted.

A few comments are in order.  First of all, we note that the values
obtained for the asymmetry coefficients, as well as the shapes of the
curves in Fig.~\ref{fig:dacp}, depend sensitively on the coefficients
$A$-$D$, $N$ and $N_3$.  We have chosen particular values for
illustration, but it is assumed that experimentalists would perform
more accurate measurements of the coefficients $A$-$D$ in tandem with
performing any CP analysis.  Also, recall that we have assumed that
$f_H$ is a constant and have thus pulled it outside of the various
integrations.  This may well be a poor approximation, in which case
the expressions for the differential CP asymmetries would need to be
modified to include the dependence that $f_H$ has on the various
variables.  Finally, we note that more recent analyses use an
expression for $F_3$ that differs from the expression we use
[Eq.~(\ref{eq:f3})].  References~\cite{finkemeiermirkes1,Grellscheid}
use an expression that is similar to Eq.~(\ref{eq:f3}), except that it
sets $\alpha=1$ and $N_3=1$, and that it replaces $BW_{K^*}(Q^2)$ and
$BW_{K^*}(s_2)$ by expressions that take into account one or both of
the $K^{*\prime}$/$K^{*\prime\prime}$ resonances.  We have performed
an analysis using this modified expression for $F_3$; the change
affects the asymmetry $A_{CP}^{(3)}$.  Retaining an overall
normalization constant and tuning it so that the Wess-Zumino
contribution still accounts for approximately $5\%$ of the
experimental branching ratio ($N_3\simeq 0.4206$), we find
$a_{CP}^{(3)}\simeq 6.8\times 10^{-5}$ and
$a_{CP\textrm{\scriptsize{,max}}}^{(3)}\simeq 7.1\times 10^{-4}$.  (We
do not quote a revised number for
$a_{CP\textrm{\scriptsize{,mod}}}^{(3)}$, since the replacement
$\cos\psi\to\left|\cos\psi\right|$ actually makes the magnitude of the
asymmetry smaller in this case.)  Comparison with
Table~\ref{tab:asymmetries} shows that the asymmetries are smaller in
magnitude in this case.  The differential plots are also affected.  We
do not consider results following from this revised expression for
$F_3$ further here, but our estimates below could easily be adapted to
take this change into account.

To determine actual CP asymmetries ($A_{CP}^{(i)}$) from the asymmetry
coefficients ($a_{CP}^{(i)}$), we need to know or be able to estimate
the quantity $f_H$Im$\left(\eta_P\right)$ [see
  Eq.~(\ref{eq:asym_form})].  Let us begin with a crude estimate by
assuming that the NP contribution to the width is ``hiding'' in the
experimental uncertainty of the branching ratio.  The experimental
branching ratio determined by BABAR is ${\cal B}(\tau^-\to
K^-\pi^-\pi^+\nu_\tau)=\left(0.273\pm
0.002\pm0.009\right)\%$~\cite{BABAR_BR}; i.e., the experimental
measurement has a relative uncertainty of approximately $3.4 \%$.  A
numerical integration of Eq.~(\ref{eq:dgam_no_asym}), performed under
the assumption that only the NP part contributes [i.e., setting
  $B_1$$=$$B_2$$=$$B_3$$=$$0$ and
  $B_4=B_4^{(2)}=\sqrt{Q^2}f_H\eta_P/m_\tau$ -- see
  Eq.~(\ref{eq:b42})], shows that the experimental uncertainty is
saturated when $\left|f_H\eta_P\right|\simeq 17.9$.  Assuming that
$\eta_P$ is purely imaginary yields upper bounds on the magnitudes of
the $A_{CP}^{(i)}$ in the range $3.9\times 10^{-4}$ to $0.012$.  Under
the same assumptions regarding $f_H\eta_P$, we also find that the
$A_{CP\textrm{\scriptsize{,max}}}^{(i)}$ range from 0.015 to 0.052.

The above estimates may be a bit optimistic, although it is difficult
to say without direct bounds on $f_H$ and $\eta_P$.  As noted in the
Introduction, the CLEO collaboration has searched for CP violation in
$\tau\to K\pi\nu_\tau$; they have set the following bound on the
scalar coupling that they denote $\Lambda$~\cite{CPVKpiexpt},
\beq
-0.172 < \mbox{Im}\left(\Lambda\right) < 0.067\,,\mbox{
  at 90\% C.L.}\, .
\eeq
The coupling $\Lambda$ is related to $\eta_S$ in
Eq.~(\ref{eq:np_hamiltonian}); $\eta_P$, however, does not receive a
direct constraint from this experiment.  $\eta_P$ should scale like
$m_W^2/m_H^2$ due to the Higgs propagator (where $m_W$ and $m_H$ are
the $W$ and charged Higgs masses, respectively).  If the Higgs has
electroweak couplings, then it would be reasonable to assume that
$\eta_P$ has a magnitude not exceeding unity.  At this point we do not
have a reliable way to estimate $f_H$.  One possibility is to infer
$f_H$ from $F_4$ using the quark equations of motion, although this
procedure may well have a large error.  As was noted above, $F_4$ for
this decay has been computed from the perspective of Chiral
Perturbation Theory in Ref.~\cite{dfm1994}.  Using the quark equations
of motion, one finds $\left| f_H\right|\sim Q^2 \left|F_4\right|/m_s$,
leading to an enhancement of $f_H$ because of the small strange quark
mass.  (This enhancement would be lost to some degree if the quark
mass were replaced by a meson mass.)  An approximate numerical
examination of $\left|F_4\right|$ derived from
Ref.~\cite{dfm1994}\footnote{We have not updated the expression to
  account for the possibility of contributions from both $K_1(1270)$
  and $K_1(1400)$.}  shows that it can be of order 1 GeV$~^{-1}$ for
some values of $Q^2$, $s_1$ and $s_2$ (it is also much smaller than
this for other values of the kinematical variables).  A crude estimate
of the maximum size of $\left|f_H\right|$ would be
$\left|f_H\right|\sim m_\tau^2\times (1~\mbox{GeV}^{-1})/m_s\sim
\left(1.777~\mbox{GeV}\right)^2\times
(1~\mbox{GeV}^{-1})/(0.095~\mbox{GeV}) \sim30$.  A more realistic
estimate for $\left| f_H\right|$ might be in the range $1$-$10$.
Combining these estimates, we see that $\left|f_H
\mbox{Im}\left(\eta_P\right)\right|$ could be of order $1$-$10$,
leading to a reduction of the possible magnitudes of the CP
asymmetries compared with our estimates above (for which we assumed
$\left|f_H \mbox{Im}\left(\eta_P\right)\right|\simeq 17.9$).

\section{Discussion and Concluding Remarks}
\label{sec:conclusions}

We have analyzed CP violation in $\tau\to K\pi\pi\nu_\tau$ due to NP
in the form of a charged Higgs boson.  Noting that the couplings of a
charged Higgs boson to the light quarks are suppressed in many models
due to the smallness of the light quarks' masses, we have observed
that CP-odd observables in $\tau\to K\pi\pi\nu_\tau$ probe
non-``standard'' NP CP violation.  An experimental search for CP
violation in $\tau\to K\pi\pi\nu_\tau$ would complement the search for
CP violation that has already taken place in $\tau\to
K\pi\nu_\tau$~\cite{CPVKpiexpt}.  In our notation, $\tau\to
K\pi\pi\nu_\tau$ is sensitive to the coupling $\eta_P$, while $\tau\to
K\pi\nu_\tau$ is sensitive to $\eta_S$.

We have analyzed four CP-odd observables in $\tau\to K\pi\pi\nu_\tau$
-- the rate asymmetry, two modified rate asymmetries and a
triple-product asymmetry.  The rate asymmetry is likely to be quite
small because it relies on the interference of the SM scalar form
factor with the NP contribution; thus, we did not make any numerical
estimates for this asymmetry.  The modified rate asymmetries and the
triple-product asymmetry result from the interference of the NP
amplitude with the SM contributions containing the form factors
$F_1-F_3$.  Adopting a particular model for the form factors and
making various assumptions, we have estimated the possible sizes of
the CP asymmetries numerically.  In our calculation it was found that
each of the asymmetries underwent a substantial cancellation upon
integration over the various phase space variables.  Experimentalists
may wish to consider differential CP asymmetries in order to avoid
some of this cancellation.  The maximal sizes of the three asymmetries
(assuming that the cancellations could be avoided by using
appropriately chosen weighting functions) were found to be in the
range 0.015 to 0.052.  These numbers were derived under the assumption
that the only constraint on the NP contribution is that it is
``hidden'' in the uncertainty of the branching ratio for $\tau\to
K\pi\pi\nu_\tau$.  The maximal magnitudes of the asymmetries decrease
if one makes more realistic assumptions regarding the hadronic form
factor $f_H$ and the NP parameter $\eta_P$.

We encourage experimentalists at the $B$ factories to analyze their
$\tau$ data sets in the manner that we have described.  Future
experiments, such as the Super $B$ factories, could provide even
greater sensitivity to these observables.

We close with a short comment regarding CP violation in $\tau^\mp\to
K^\mp \pi^\mp K^\pm\nu_\tau$.  In principle, this decay mode could be
analyzed in a similar manner to what we have described.  (See
Ref.~\cite{kilian}, for example.)  One advantage of $\tau\to K\pi
K\nu_\tau$ is that there is an intermediate pseudoscalar resonance
[the $\pi^\prime(1300)$] that could potentially enhance the hadronic
current associated with the NP charged scalar exchange.  We wish to
point out what appears to be an error, or an oversimplification, in
the literature regarding this point.  The scalar form factors
associated with the $\pi^\prime$ resonance in the $\tau\to
3\pi\nu_\tau$ and $\tau\to K \pi K\nu_\tau$ decays have been written
down in Ref.~\cite{dmsw}.  The expression for the $3\pi$ case seems
to be sensible, but the one for the $K\pi K$ case appears to make an
unphysical assumption regarding the contributing decay chains.  In
particular, judging from the expression, one of the decay chains would
seem to have an intermediate $\rho$ decaying to a $K$ and a $\pi$.  If
this is remedied by replacing the $\rho$ by a $K^*$, one finds that
none of the decay chains can quite proceed on shell (although there is
a large uncertainty in the $\pi^\prime$ mass; furthermore, the
$\pi^\prime$ does have a large width and the decay $\pi^\prime \to K^*
K$ is actually right near threshold).

\bigskip
\noindent
{\bf Acknowledgments}:
We thank the following people for helpful correspondence and
discussion: M. DeLong, H. Hayashii, I. Kravchenko, P. Richardson,
M. Roney and A. Weinstein.  K.K. thanks the Physics Department of the
Universit\'e de Montr\'{e}al for its hospitality while part of this
work was performed. This work was financially supported by NSERC of
Canada. The work of K.K. and K.L. was supported in part by the
U.S.\ National Science Foundation under Grant PHY--0601103.


\end{document}